\documentclass{ptapap}
\author{Dawid Mo\'zdzierski}[IAUWr]
\author{Andrzej Pigulski}[IAUWr]
\affil[IAUWr]{Astronomical Institute, University of Wroc{\l}aw,\\ Kopernika 11, 51-622 Wroc{\l}aw, Poland}

\title{Prospects for the ensemble asteroseismology in
young open clusters}

\begin{document}

\maketitle

\begin{abstract}

This is a progress report on the ongoing project dealing with ensemble asteroseismology of B-type
stars in young open clusters. The project is aimed at searches for B-type pulsating stars in open clusters,
determination of atmospheric parameters for some members and seismic modeling of B-type pulsators. Some
results for NGC\,457, IC\,1805, IC\,4996, NGC\,6910 and $\alpha$ Per open clusters are presented. For the last
cluster, BRITE data for five members were used.

\end{abstract}

\section{Introduction}

Stellar clusters are known as good `laboratories' for studying member stars. By fitting isochrones,
one can obtain distance, age and sometimes get information on metallicity of a cluster. Our project is
focused on finding open clusters rich in pulsating B-type stars and subsequent seismic modeling of their members by means of 
ensemble asteroseismology (hereafter EnsA). EnsA takes advantage of the common parameters of the members of a cluster (e.g.~age and metallicity) 
to put additional constraints on seismic models of member stars. It is applicable to open clusters rich in massive pulsating stars. 
The most promising are $\beta$ Cep stars for which many interesting results 
were already obtained by means of seismic modeling \citep[e.g.][]{Aerts2003, Pam2004, Dup2004, Dasz2010}.
One of the best candidates for application of EnsA is NGC\,6910 \citep{Kol2004}, for
which pulsation parameters were obtained as a result of the international observational campaign
\citep{Pig2008, Sae2010}. In the era of nano-satellites, new possibilities of seismic
studies of stars belonging to bright, nearby star clusters occured. An example is $\alpha$ Per open cluster.

\begin{figure}
\begin{center}
\includegraphics[width=9cm]{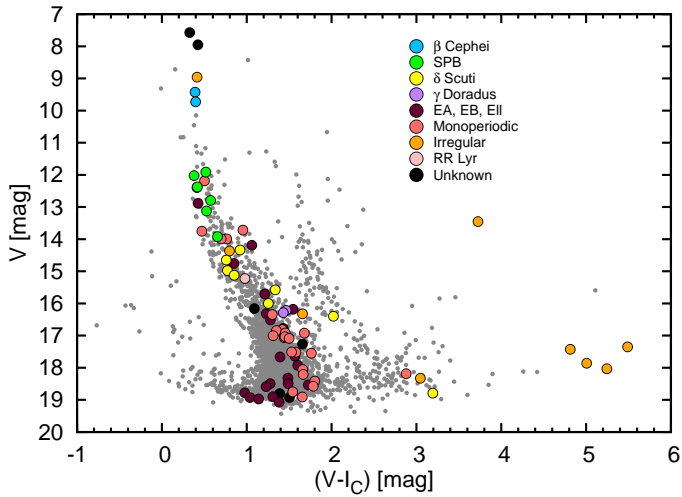}
\caption{Variable stars in the color-magnitude diagram of IC\,4996.}
\label{VarCMD}
\end{center}
\end{figure}

\section{New observations and results}
The results of the observations of NGC\,457 and preliminary results of the search for variable stars in IC\,1805 were published
by \cite{Moz2014} and \cite{Moz2012}, respectively. Recently, we performed also search for variable stars in another young open cluster, IC\,4996. Observations
of IC\,4996 were made between 2007 and 2014 in Bia{\l}k\'ow Observatory (Poland) during 50 observing nights. The observations were carried out with a 60-cm
reflecting telescope with the attached CCD camera covering $13^\prime\times12^\prime$ field of view. About 7500 CCD frames were acquired through
the $B$, $V$, $R$, $I_{\rm C}$ and narrow-band H$\alpha$ filters. We detected 81 variable stars (Fig. \ref{VarCMD}), of which 71 are new. One new $\beta$ Cep star
was found.

In total, in three open clusters  (NGC\,457, IC\,1805, and IC\,4996) we have detected 231 variables. Among the most interesting are: large
population of SPB stars (21) in NGC\,457, some of them showing frequences above 3.5 $\rm d^{-1}$, and many
monoperiodic variables in IC\,4996 and IC\,1805 grouping in the lower parts of their color-magnitude diagrams.
These are probably pre-main sequence stars. 
Small number of $\beta$ Cep stars in NGC\,457, IC\,4996 and IC\,1805 does not give, or gives marginal
chance for a successful application of EnsA. Nonetheless, our study resulted in finding mentioned above groups of SPB and monoperiodic stars which in
the future might be useful for better understanding of the incidence of variability in young open clusters at main sequence and 
pre-main sequence stages of evolution. 

The best candidate for the application of EnsA is NGC\,6910. The first results look promising.
Using echelle spectra obtained with Apache Point Observatory (APO) ARC
3.5-m telescope and Nordic Optical Telescope (NOT), we have derived atmospheric
parameters of three $\beta$ Cep stars from NGC\,6910, NGC\,6910-14, -16 and -18, using NLTE BSTAR2006 grid \citep{Lanz2007} of atmosperic models.
Then, effective temperatures and surface gravities of these stars were used to place them
in the theoretical H-R diagram and for mode identification. In calculations, we used Warsaw - New Jersey evolutionary code adopting
OPAL opacities, solar mixture as determined by \citet{Asp2009} and no overshooting from the convective core. Mode identification
based on $B$, $V$, $I_{\rm C}$ time-series photometry was performed with the methods developed by \cite{Dasz2003} and \cite{Dasz2005}. We identified degree
of the mode with frequency $f=5.252056$ $\rm d^{-1}$ detected in NGC\,6910-14 as $l$ = 4. In view of the relatively high projected rotational velocity
of the star ($V_{\rm rot} = 149$ km/s), the possiblity that it is rotationally coupled $l=$ 2 mode, cannot be excluded. 
We also identified degrees of four modes with the highest photometric amplitudes in two other $\beta$~Cephei stars, NGC\,6910-16
($f_1=5.202740$~$\rm d^{-1}$ and $f_2=4.174670$ $\rm d^{-1}$, both as $l=$ 2) and NGC\,6910-18 ($f_1=6.154885$ $\rm d^{-1}$ as $l$ = 0 and $f_2=6.388421$
$\rm d^{-1}$ as $l$ = 2). The full results of the application of EnsA to NGC\,6910 will be published elsewhere.

Unfortunately, it turned out that the up-to-date BRITE observations of five B-type members
of $\alpha$ Per cluster revealed only one pulsating star, HD\,22192. It can be classified as an SPB star with frequency groupings.  Therefore, the cluster does not
seem to be an object suitable for the application of EnsA.

\acknowledgements{This work was supported by the NCN grant No. 2012/05/N/ST9/03898 and has received funding from the European Community's Seventh
Framework Programme (FP7/2007-2013) under grant agreement no. 269194. We thank Z. Ko{\l}aczkowski, D. Drobek,
A. Majewska, M. St\k{e}\'slicki, G. Kopacki, E. Niemczura and G. Michalska for making observations of IC\,4996 in Bia{\l}k\'ow.
We thank Dr. J. Jackiewicz for making available telescope time on the APO ARC 3.5-m telescope and for his support during
observations. We also thank for T. Morel and J.H. Telting for making spectroscopic observations on NOT and P.~Walczak
for providing some software and patient comments. This research has made use of the WEBDA database, operated at the Department
of Theoretical Physics and Astrophysics of the Masaryk University. The work on $\alpha$~Per is based on data collected by the BRITE-Constellation satellite mission,
built, launched and operated thanks to support from the Austrian
Aeronautics and Space Agency and the University of Vienna, the Canadian
Space Agency (CSA) and the Foundation for Polish Science \&
Technology (FNiTP MNiSW) and National Centre for Science (NCN).
}

\bibliographystyle{ptapap}
\bibliography{prospects}

\end{document}